\documentclass{ws-procs9x6}

\begin{document}

\title{INCLUSIVE-JET AND DIJET CROSS-SECTIONS IN HIGH-$Q^2$ DIS AT HERA}

\author{T.~SCH\"ORNER-SADENIUS\footnote{
    \uppercase{T}alk given on
    behalf of the \uppercase{ZEUS} collaboration at \uppercase{DIS}06,
    \uppercase{T}sukuba, \uppercase{J}apan, \uppercase{A}ril 2006.}
%    \uppercase{O}n behalf of the \uppercase{ZEUS} collaboration.}
}

\address{Hamburg University, IExpPh, \\
Luruper Chausse 149, 22761 Hamburg, Germany \\  
E-mail: schorner@mail.desy.de}

\maketitle

\abstracts{
A new ZEUS measurement of inclusive-jet and dijet cross-sections in the Breit
frame performed in deep-inelastic ep scattering data is presented. 
The data correspond to an integrated luminosity of about 82~${\rm
  pb^{-1}}$ and are 
resctricted to the kinematic regime $Q^2 >$~125~${{\rm GeV^2}}$ and -0.65~$<
\cos\gamma_{had} <$~0.65, where $Q^2$ is the photon virtuality and
$\cos\gamma_{had}$ corresponds to the polar angle of the hadronic
system. The cross-sections are measured as functions of various
kinematic and jet observables and are compared to NLO QCD calculations which
describe the data well within all uncertainties. 
}

\section{Introduction}

Jet measurements provide stringent tests of the concepts of perturbative QCD
and factorisation and offer access to the central parameter of QCD, the strong
coupling constant $\alpha_S$. 
In addition, jet cross-sections as measured in
ep collisions at HERA are sensitive to the proton parton distribution
functions (PDFs). This fact has been exploited recently by the ZEUS
collaboration who 
included jet cross-section measurements from both photoproduction and
deep-inelastic scattering (DIS) into their NLO QCD fits for the
PDFs\cite{zeusfit}. 

In this contribution a new measurement of inclusive-jet and dijet
cross-sections in DIS at high values of the negative squared four-momentum of
the exchanged boson, $Q^2 >$~125~${\rm GeV^2}$, is presented. The data used in
the analysis extend previous analyses both in statistics, considering almost
three times 
that of an earlier analysis of inclusive-jet cross-sections at
ZEUS\cite{oscar}, and in kinematic range (proton energy $E_p$ of 920~GeV
instead of 820~GeV).

For both the inclusive-jet and dijet measurement the data are presented as
double-differential cross-sections in $Q^2$ and in the jet transverse energy
in 
the Breit frame, $E_{T}$ (for the inclusive-jet analysis) and in $Q^2$
and $\xi$ (for the dijet analysis, $\xi\equiv x_{Bj}\cdot\left(
1+M^2_{jj}/Q^2\right)$ is the momentum fraction carried by the struck
parton). These observables provide optimal sensitivity to the parton
distributions functions and might therefore serve as input to future QCD fits
of the proton PDFs. 

\section{Data Selection, Correction and Theoretical Predictions}

The data used in the analysis were collected with the ZEUS detector at HERA in
the years 1998-2000 and correspond to an integrated luminosity of 81.7~${\rm
  pb^{-1}}$. The phase space of the analysis is defined by the following two
requirements: $Q^2 >$~125~${\rm GeV^2}$ and 
-0.65~$< \cos\gamma_{had} <$~0.65. 
Here, $\cos\gamma_{had}$ is the polar
angle of the hadronic system which, in events of the Quark-Parton Model
type, corresponds to the angle of the struck quark.
In case of the
dijet analysis $Q^2 <$~5000~${\rm GeV^2}$ was required in addition 
in order to restrict the data
to a regime where the $Z^0$ exchange can savely be neglected. 
Jet reconstruction is performed using 
the longitudinally invariant $k_T$ algorithm\cite{kt} in the inclusive
mode\cite{jetalgo} in the Breit frame. 
Jets were then selected requiring their pseudorapidity to be in the 
interval -2~$< \eta_{Breit} <$~1.5 
and to have transverse jet energies of at least 8~GeV (inclusive-jet 
analysis) or 12~GeV and 8~GeV (dijet analysis).
 
The data were corrected for detector efficiency and acceptance effects using
the {\sc Lepto}\cite{lepto} (dijet analysis) or {\sc Ariadne}\cite{ariadne} 
Monte Carlo (MC) models (inclusive-jet analysis). Also higher-order QED
effects were taken into account using these MC models. 

The NLO QCD calculations used for the comparison with the data 
were performed with the {\sc Disent}
program\cite{disent} using the latest CTEQ6 proton PDFs\cite{cteq6}. Since the
calculations provide jets of partons whereas the corrected data correspond to
jets of hadrons, the NLO QCD calculations were corrected to the hadron level
using the average of the corrections predicted by the {\sc Lepto} and {\sc
  Ariadne} models. The inclusive-jet predictions were in addition corrected
for the effects of the $Z^0$ exchange.   

\section{Systematic Checks}

On the theory side, the uncertainty of the hadronisation correction, the
uncertainties due to the uncertain knowlegde of the strong coupling parameter
$\alpha_S$ and on the input proton PDFs and the uncertainty due to neglected
higher orders in the perturbative expansion were considered. The last
contribution, which was estimated by a variation of the renormalisation scale
by a factor 2 up and down, is dominating, with effects on the predicted
cross-sections of up to 20$\%$ in certain kinematic regions. 
In a few regions, especially at high $Q^2$, also the PDF uncertainty
gets significant, reaching up to 4$\%$.  

The main experimental uncertainty comes from the uncertainty on the hadronic
energy scale. This uncertainty, which is assumed to be the only correlated
uncertainty leads to effects on the cross-sections of typically 10$\%$.

\section{Results}

\begin{figure}[t]
%\vfill
%\setlength{\unitlength}{1.0cm}
%\begin{picture}
\hspace*{-0.5cm}
\epsfxsize=6.5cm   %width of figure - will enlarge/reduce the figures
\epsfbox{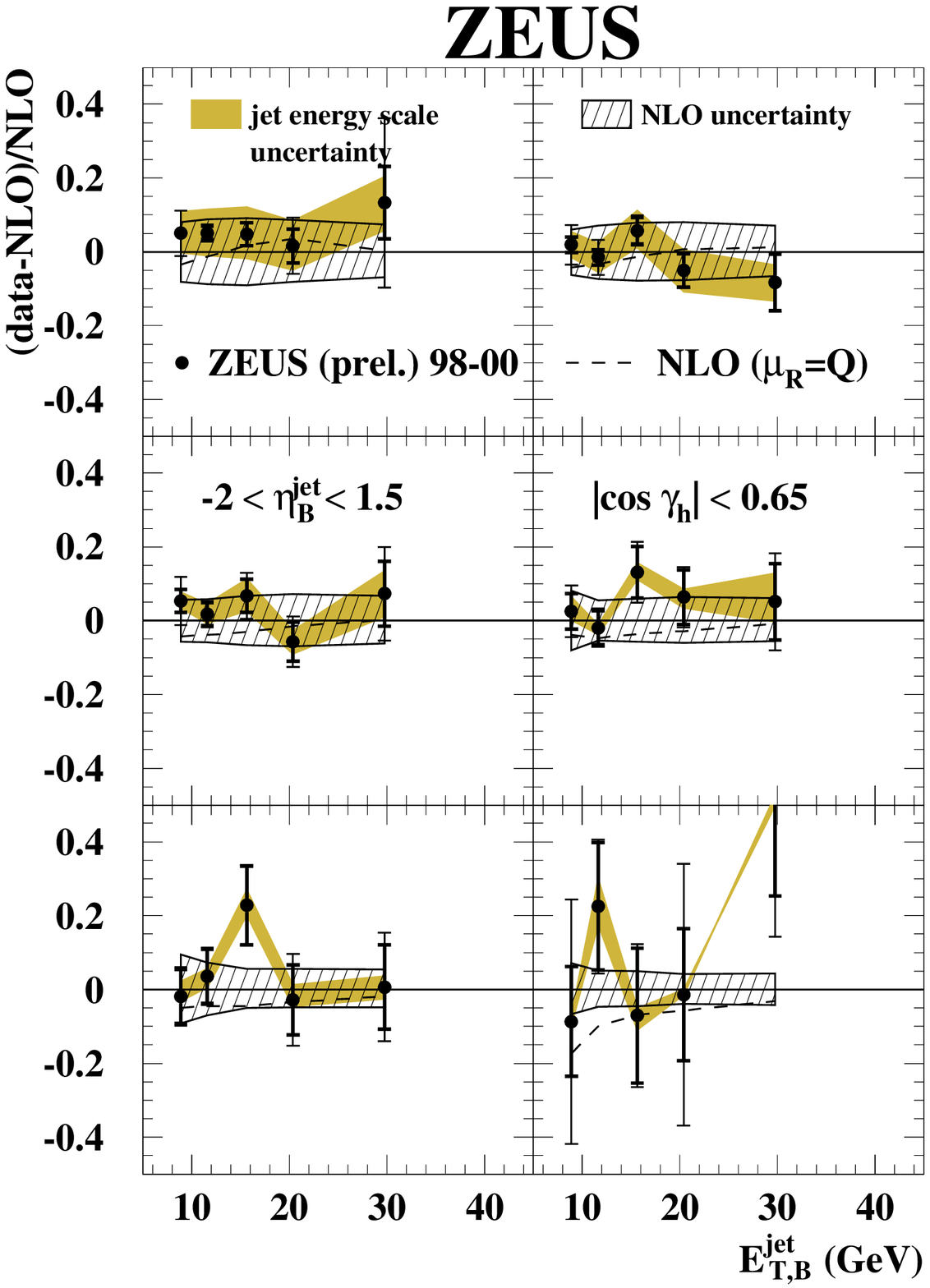}
\hspace*{-1.cm}
\epsfxsize=6.cm   %width of figure - will enlarge/reduce the figures
\epsfbox{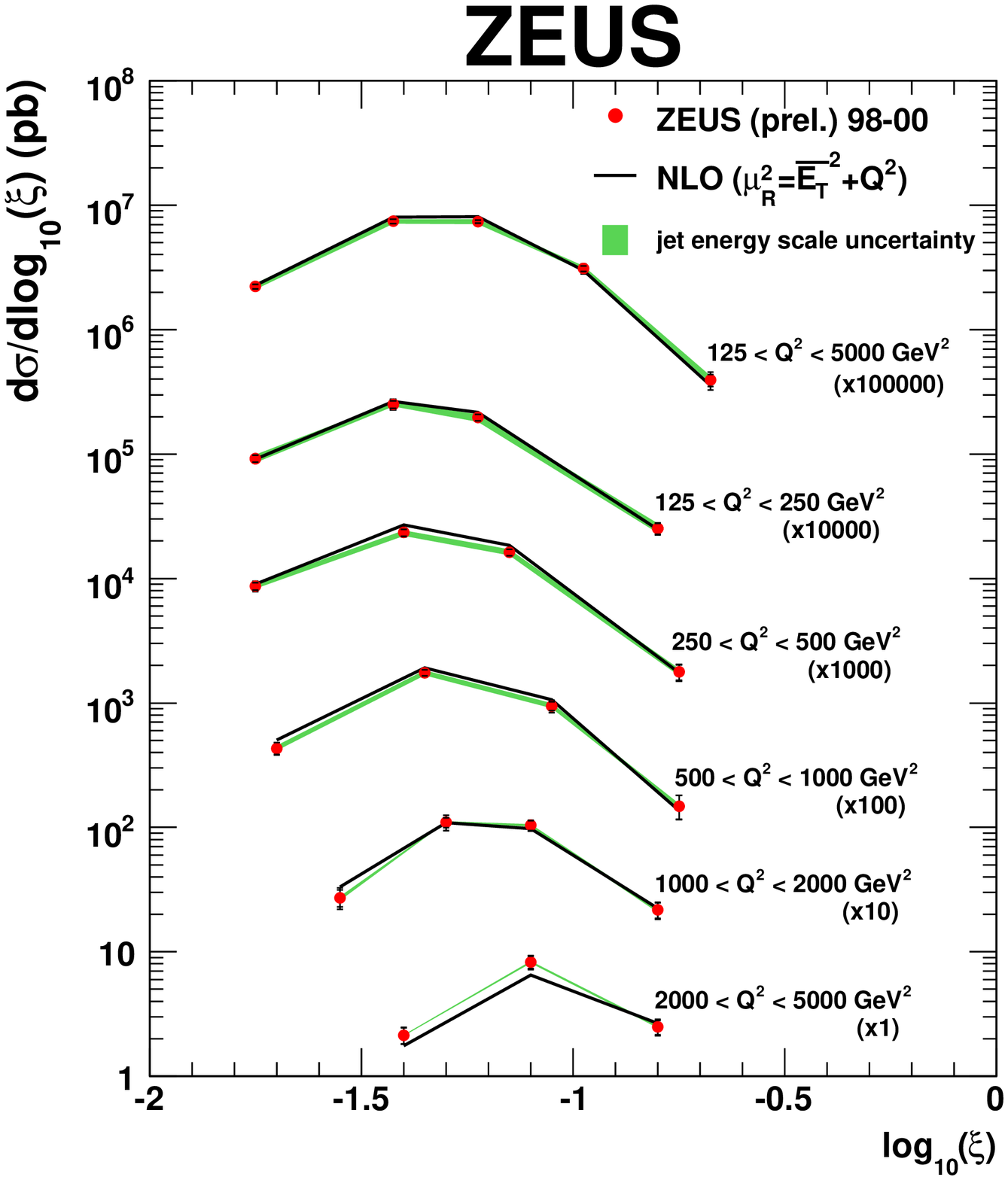}
%\epsfbox{figure2.eps}
%\figurebox{2cm}{3cm}{} %to have a box alone 
%\centerline{\epsfxsize=4.1in\epsfbox{procs-fig1.eps}}   
%\put (1.0,0.0){\epsfig{figure= figure1.eps,width=5.cm}}
%\end{picture}
\caption{Left: Ratio $(data-NLO)/NLO$ for the inclusive-jet analysis in 6 different
  regions of $Q^2$ as functions of $E_T$. Right: Double-differential dijet
  cross-sections in different regions of $Q^2$ as functions of $\log\xi$. The
  data points are compared to the NLO QCD calculations. See text for more
  details. \label{figure1}}
\end{figure}

Inclusive-jet cross-sections were measured double-differentially in regions of
$Q^2$ as functions of $E_T$. The data are well described by the NLO QCD
calculations within all uncertainties. Figure~\ref{figure1} (left) shows, in 6
different regions of $Q^2$, the ratio $(data-NLO)/NLO$ together with
statistical and uncorrelated systematic uncertainties. The shaded band
indicates the correlated systematic uncertainty due to the jet energy scale,
and the hashed area gives the theoretical uncertainty. 

The dijet cross-sections were measured single-inclusively in various kinematic
and dijet observables like $Q^2$, invariant dijet mass $M_{jj}$ or $\log\xi$
and are in general well described by the QCD calculations. In addition,
double-differential dijet cross-sections as functions of $\log\xi$ in 5
different regions of $Q^2$ were measured. Also these distributions are well
reproduced by the theoretical predictions, although at high values of $Q^2$
statistical effects start to dominate. Figure~\ref{figure1} (right) shows the
measured double-differential cross-sections together with the NLO
predictions. 

%\begin{figure}[ht]
%%\vfill
%%\setlength{\unitlength}{1.0cm}
%%\begin{picture}
%\hspace*{1.5cm}
%\epsfxsize=8.cm   %width of figure - will enlarge/reduce the figures
%\epsfbox{figure3.eps}
%%\epsfbox{figure2.eps}
%%\figurebox{2cm}{3cm}{} %to have a box alone 
%%\centerline{\epsfxsize=4.1in\epsfbox{procs-fig1.eps}}   
%%\put (1.0,0.0){\epsfig{figure= figure1.eps,width=5.cm}}
%%\end{picture}
%\vspace*{-1.cm}
%\caption{Contribution of gluon-induced events to the inclusive-jet and dijet
%  cross-sections as functions of $Q^2$. See text for more details. \label{figure2}}
%\end{figure}

The contribution of gluon-induced events to the total cross-section was also
estimated
as a function of $Q^2$ separately for both the inclusive-jet analysis
and for the dijet analysis. 
It was observed  that even for very high $Q^2$ values around
3000~${\rm GeV^2}$ there is still a substantial contribution from
gluon-induced processes of about 30$\%$ for the dijet case and about 20$\%$
for the inclusive-jet analysis. The observed theoretical 
uncertainties due to the uncertainty in the gluon density were sizeable; 
therefore these data promise further constraints on the gluon density when included in QCD fits.   

\section{Conclusion and Outlook}

Dijet cross-sections at high values of $Q^2 >$~125~${\rm GeV^2}$ have been
measured single- and double-differentially in various kinematic and dijet
quantities. In addition, inclusive-jet cross-sections were studied
double-differentially in bins of $Q^2$ as functions of the jet transverse
energy. 

The very precise data extend former ZEUS measurements of inclusive-jet and
dijet cross-sections in high-$Q^2$ DIS. They are well described by the NLO QCD
calculations and provide access to 
the parton distribution functions of the proton, especially the gluon density
at high values of the proton momentum fraction. Therefore, the
measurements are natural candidates for future use in NLO QCD fits of the
parton densities.

\end{document}